# Transitivity of Commutativity for Linear Time-Varying Analog Systems

Mehmet Emir KOKSAL

*Department of Mathematics, Ondokuz Mayis University, 55139 Atakum, Samsun, Turkey*

emir_koksal@hotmail.com

**Abstract:** In this contribution, the transitivity property of commutative first-order linear time-varying systems is investigated with and without initial conditions. It is proven that transitivity property of first-order systems holds with and without initial conditions. On the base of impulse response function, transitivity of commutation property is formulated for any triplet of commutative linear time-varying relaxed systems. Transitivity proves are given for some special combinations of first and second-order linear time-varying systems which are initially relaxed.

1.      Introduction

As the main branches of applied mathematics, differential (and integral) equations arise in many areas of sciences and engineering including acoustics, electromagnetic, electrodynamics, fluid dynamics etc. There is a great deal of papers on the theory, technique and applications of differential equations. Especially, they are used as a major tool in order to achieve many developments in real engineering problems by modelling, analyzing and solving naturel problems. For example, an interdisciplinary branch of applied mathematics and electric-electronics engineering, they play a pioneering role in system and control theory, that deal with the behavior of dynamical systems with inputs, and how their behavior is modified by different combinations such as cascade and feedback connections which is. When the cascade connection in system design is considered, the commutativity concept places a prominent role to improve different system performances.

Consider a system $A$ described by a linear time-varying differential equation of the form

$$A: \sum_{i=0}^{n_A} a_i(t)\frac{d^i}{dt^i}y_A(t) = x_A(t); \qquad (1)$$

$(x - y)$ represents input-output pair of the system at any time $t \in R$, $a_i(t)$ are time-varying coefficients with $a_n(t) \neq 0$, $n_A \geq 0$ is the oder of system; and $y_A^{(i)}(t_0) \in R$, $i = 0, 1, \cdots, n_A - 1$ are the initial conditions at the initial time $t_0 \in R$.

When two systems of this type are interconnected sequentially so that the output of the former feeds the input of the later, it is said that they are connected in cascade [1]. If the order of connection does not affect the input-output relation of the combined system $AB$ or $BA$, it is said that systems $A$ and $B$ are commutative.

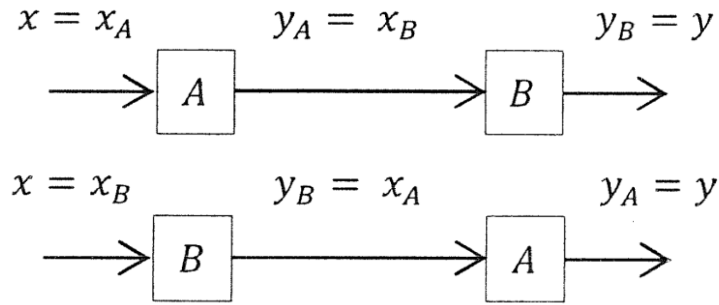

**Figure 1:** Cascade connection of the differential system $A$ and $B$

If the combined system has an overall input-output relation invariant with the sequence of connection, it is said that these systems are commutative [2]. In [2], J. E. Marshall has proven that "for commutativity, either both systems are time-invariant or both systems are time-varying". In addition, he has revealed the commutativity conditions of first-order systems.

Later Marshall's work, a great deal of researches has been done on commutativity. In [3-5], the necessary and sufficient conditions for commutativity of second-order systems are presented. Koksal has presented the general commutativity conditions for time-varying systems of any order and reformulated the previous results obtained for seccond-order systems in the format of general conditions [6]. In this work, the general conditions are used to show that any system with constant forward and feedback gains is commutative with the system itself, which is an important fact for the feedback control theory.

In 1985, Koksal prepared a technical report which is a survey on commutativity [7]. In this report, an iterative formula is derived and an explicit formula is given for the entries of the coefficient matrix expressing the first set of commutativity conditions and the second set of commutativity conditions, respectively. Hence, the theorem stating these conditions is formally proved. Morover, explicit commutativity results for fourth-order systems are obtained. Finally, commutativity of Euler's system is proved.

The content of the published but undistributed work [7] can be found in the exhaustive journal paper of M. Koksal introduced the basic fundamentals of the subject [8]. His paper covers almost all the previous results except the ones related with initial conditions and sensitivity. This paper is the first tutorial paper that has appeared in the literature.

More than one decade no publication had been appeared in the literature until the work in 2011 [9]. This reference is the second basic journal publication after the first appeared in 1988 [8]. In [9], another generic paper by the same author has presented explicit commutativity conditions of fifth-order systems in addition to reviews of commutativity of systems with non-zero initial conditions, commutativity and system disturbance, commutativity of Euler systems.

Research on commutativity has not confined to analog systems only; there has been some literature on the subject for discrete-time systems as well [10, 11]. Hence, the research is continuing on both digital and analog area [12]. In [12], explicit results for finding all the second order commutative pairs of a first-order linear time-varying system have been given and the derived theoretical results have been verified by an example. About the commutativity of continuous time linear time-varying systems, [13] has been the last paper appearing in the literature; it deals with driving necessary and sufficiently conditions for the decomposition of a second order linear time-varying system into two cascade connected commutative first-order linear time-varying subsystems.

After a short introduction of the literature in this section, the transitivity property of commutativity is introduced in Section 2. Transitivity property of commutative first-order systems with and without initial conditions is studied in Section 3. In Section 4, transitivity property of commutativity for relaxed systems of any order is formulated on the base of impulse response function. Section 5 covers the verification of the general results formulated in Section 4 on the base of impulse response function for first-order systems. Section 6 illustrates the results presented in Sections 4 and 5. Section 7 includes transitivity proves for some combinations of first and second-order relaxed linear time-varying systems. Finally, the paper finishes with conclusions and future work which compose the last section, Section 8.

## 2. Transitivity Property of Commutativity

Logically or mathematically, transitivity is a property of a binary relation such that whenever one element is related to a second element, and the second element is related to a third element, then the first element is also related to the third element.

Let $A, B, C$ be dynamical systems described by linear time-varying differential equations of the form (1) with their special input-output variables, orders of complexity, and coefficients. If $A$ and $B$ are commutative among themselves, further, $B$ and $C$ are also commutative amoung themselves, what can it be said about the commutativity of systems $A$ and $C$. If $A$ and $C$ are also commutative among themselves, this property is called transitivity property of commutativity, in other words commutativity is a transitive relation.

For commutative systems $A, B, C$ satisfiying transitivity, all the triplets $ABC, ACB, BAC, BCA, CAB, CBA$ yield the same functionally equivalent cascade connected system and hence the same input-output property. The preference of an individual system connection depends on its relative performance characteristics such as sensitivity, disturbance, robustness, and etc., with respect to the others.

It is well-known (or trivial to show) that linear scalar systems, that is systems of order $0$, are always commutative among themselves whether time-invariant or time-varying; and transitivity always hold for such systems. Further, initially relaxed time-invariant systems are always commutative. When initial conditions exist, which is valid for non-scalar (1st or higher-order) systems, the commutativity of such systems (even including a scalar system) is not automatic and requires some conditions. Hence, the following discussions are devoted mainly to commutativity of systems at least one of them is of order one or higher.

## 3. Transitivity for First-order Systems

Let $A, B, C$ be first-order linear time-varying systems defined by a differential equation of the form (1). Let $a_1, a_0$; $b_1, b_0$; and $c_1, c_0$ are time-varying (in general) coefficients of these systems. Assume also that $(A, B)$ and $(B, C)$ are commutative pairs.

The first commutativity conditions [7-9] for $(A, B)$ and $(B, C)$ yields

$$\begin{bmatrix} b_1 \\ b_0 \end{bmatrix} = \begin{bmatrix} a_1 & 0 \\ a_0 & 1 \end{bmatrix} \begin{bmatrix} k_1 \\ k_0 \end{bmatrix}, \qquad (2a)$$

$$\begin{bmatrix} c_1 \\ c_0 \end{bmatrix} = \begin{bmatrix} b_1 & 0 \\ b_0 & 1 \end{bmatrix} \begin{bmatrix} \ell_1 \\ \ell_0 \end{bmatrix}, \tag{2b}$$

respectively; where $k_1, k_0, \ell_1, \ell_0$ are some constants.

Inserting the values of $b_1, b_0$ as expressed in (2a) into Eq. (2b) and rearranging, we obtain

$$\begin{bmatrix} c_1 \\ c_0 \end{bmatrix} = \begin{bmatrix} a_1 & 0 \\ a_0 & 1 \end{bmatrix} \begin{bmatrix} k_1 \ell_1 \\ k_0 \ell_1 + \ell_0 \end{bmatrix}. \tag{2c}$$

This equation already satisfies the first commutativity condition for Systems $A$ and $C$; hence, $A$ and $C$ are commutative pairs under zero initial conditions.

For the case of existence of non-zero initial conditions, the second commutativity condition [9] for $(A, B)$ and $(B, C)$ yields

$$y_A(t_0) = y_B(t_0) \neq 0, \tag{3a}$$

$$\frac{1 - a_0(t_0)}{a_1(t_0)} = \frac{1 - b_0(t_0)}{b_1(t_0)}; \tag{3b}$$

$$y_B(t_0) = y_C(t_0) \neq 0, \tag{4a}$$

$$\frac{1 - b_0(t_0)}{b_1(t_0)} = \frac{1 - c_0(t_0)}{c_1(t_0)}, \tag{4b}$$

respectively. Hence, Eqs. (3) and (4) lead to write

$$y_A(t_0) = y_C(t_0) \neq 0, \tag{5a}$$

$$\frac{1 - a_0(t_0)}{a_1(t_0)} = \frac{1 - c_0(t_0)}{c_1(t_0)}. \tag{5b}$$

The result in Eq. (5) clearly shows the satisfaction of the second commutativity condition for Subsystems $A$ and $C$. Hence, $A$ and $C$ are also commutative when initial conditions exist, and transitivity property is valid for non-relaxed first-order linear time-varying systems as well.

In fact, Eq. (3b) is not satisfied for all systems $A$ and $B$ satisfying commutativity conditions (2a) under relaxed conditions. Using (2a) we see that Eq. (3b) requires

$$\frac{1 - a_0(t_0)}{a_1(t_0)} = \frac{1 - b_0(t_0)}{b_1(t_0)} = \frac{1 - k_1 a_0(t_0) - k_0}{k_1 a_1(t_0)} = \frac{\frac{1 - k_0}{k_1} - a_0(t_0)}{a_1(t_0)}; \tag{6a}$$

Or equivalently

$$k_0 = 1 - k_1. \tag{6b}$$

Similarly, using (2b), we see that Eq. (4b) requires

$$\frac{1 - b_0(t_0)}{b_1(t_0)} = \frac{1 - c_0(t_0)}{c_1(t_0)} = \frac{1 - \ell_1 b_0(t_0) - \ell_0}{\ell_1 b_1(t_0)} = \frac{\frac{1-\ell_0}{\ell_1} - b_0(t_0)}{b_1(t_0)}, \tag{7a}$$

$$\ell_0 = 1 - \ell_1. \tag{7b}$$

Using (2c) with (6b) and (7b), we can easily show that (5b) is satisfied; That is:

$$\frac{1 - c_0(t_0)}{c_1(t_0)} = \frac{1 - [k_1 \ell_1 a_0(t_0) + k_0 \ell_1 + \ell_0]}{k_1 \ell_1 a_1(t_0)} = \frac{\frac{1 - k_0 \ell_1 - \ell_0}{k_1 \ell_1} - a_0(t_0)}{a_1(t_0)}$$

$$= \frac{\frac{1 - (1 - k_1)\ell_1 - (1 - \ell_1)}{k_1 \ell_1} - a_0(t_0)}{a_1(t_0)} = \frac{1 - a_0(t_0)}{a_1(t_0)}, \tag{8}$$

which is exactly Eq. (5b); hence, $A$ and $C$ are commutative under non-zero conditions as well if the constants $k_1, \ell_1, k_0, , \ell_0$ satisfy (6b) and (7b). These are explicit commutativity conditions under non-zero initial states equivalently replacing by (3b) and (4b) in addition to (2a) and (2b) which are necessary and sufficient for relaxed case. With these conditions 5b is also satisfied, so that transitivity is valid.

We state the results obtained so far by two theorems:

***Theorem 1:*** *The necessary and sufficient conditions that two first-order linear time-varying systems A and B described by differential equations of the form (1) are commutative under zero initial conditions are that*

i. *The coefficients of System B must be expressible in terms of the coefficients of those of A as in Eq. (2a) where $k_1 \neq 0$ and $k_0$ are arbitrary constants.*

ii. *Further, Systems A and B are commutative under arbitrary non-zero initial conditions as well if and only if Condition (i) is satisfied with $k_0 = 1 - k_1$ (that is $k_1 \neq 0$ and $k_0$ cannot be chosen arbitrarily), and their non-zero initial conditions must be equal.*

Note that this theorem is the first theorem expressing the commutativity conditions for unrelaxed systems by using the arbitrary constants relating the coefficients of two systems for their commutativity in relaxed conditions. Note also that Eqs. (6b) and (7b) are independent of the initial time. So, commutativity does not depend on the initial time $t_0$ for first-order non-relaxed systems.

***Theorem 2:*** The transitivity property of commutativity for first-order linear time-varying systems is always valid for both relaxed and un-relaxed commutative systems that is for systems with zero and non-zero initial conditions.

## 4. Transitivity for High-order Systems

In this section, higher-order linear time-varying systems with zero initial conditions are considered; scalar (0-order) and first-order systems are also included in the scope. To prove the transitivity property of commutativity, the impulse response function is used. The results are general in the sense the subsystems of each commutative pair may have a different order than its partner.

Let $A$ be a liear time-varying initially relaxed system having an impulse response $h_A(t,\tau)$; by definition this is the response to a unit impulse occurring at time $\tau$ and observed at time $t \geq \tau$ [14]. And the response of $A$ for an arbitrary input $u_A(t)$ is expressed by superposition integral

$$y_A(t) = \int_{t_0}^{t} h_A(t,\gamma) u_A(\gamma) d\gamma. \qquad (9a)$$

Consider now the cascade connection of $A$ and $B$ as shown in Fig. 2 and denote this connection by $AB$. Smilar to Eq. (9a), for the output of $B$ we write

$$y_B(t) = \int_{t_0}^{t} h_B(t,\gamma) u_B(\gamma) d\gamma. \qquad (9b)$$

It is obvious form Fig. 1, and using Eq. (9a)

$$u_B(\tau) = \int_{t_0}^{\tau} h_B(\tau,\gamma) u(\gamma) d\gamma. \qquad (9c)$$

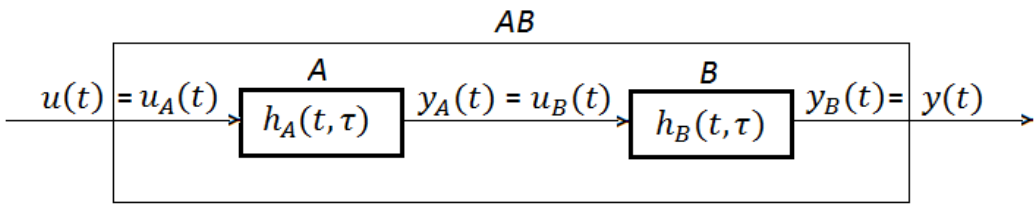

**Figure 2:** Cascade connection $AB$ of Systems $A$ and $B$.

Finally, inserting Eq. (9c) in Eq. (9b), and observing $y_B(t)$ in Fig. 2, we obtain

$$y(t) = \int_{t_0}^{t} h_B(t,\tau)\left[\int_{t_0}^{\tau} h_A(\tau,\gamma)u(\gamma)d\gamma\right]d\tau. \tag{9d}$$

This is the response of $AB$ for any input $u(t)$ applied for $t \geq t_0$. To find the impulse response $h_A(t, t_0)$ of the connection $AB$, we substitute $u(t) = \delta(t - t_0)$ in Eq. (9d) and arrive

$$h_{AB}(t,t_0) = \int_{t_0}^{t} h_B(t,\tau)\left[\int_{t_0}^{\tau} h_A(\tau,\gamma)\delta(\gamma - t_0)d\gamma\right]d\tau = \int_{t_0}^{t} h_B(t,\tau)h_A(\tau,t_0)d\tau, \tag{10a}$$

where the second equality results from the property of impulse function. In a similar result, the impulse response $h_{BA}(t,\tau)$ of the connection $AB$ can be written as

$$h_{BA}(t,t_0) = \int_{t_0}^{t} h_A(t,\tau)h_B(\tau,t_0)d\tau. \tag{10b}$$

Since equal impulse responses yield equal input-output pairs, for commutativity of $A$ and $B$ it must be true that

$$h_{AB}(t,t_0) = h_{BA}(t,t_0); \quad \forall t \geq t_0, \forall t_0 \in R. \tag{11a}$$

Using Eqs. (10a, b); (11a) can be written as

$$\int_{t_0}^{t} h_B(t,\tau)h_A(\tau,t_0)d\tau = \int_{t_0}^{t} h_A(t,\tau)h_B(\tau,t_0)d\tau; \quad \forall t \geq t_0, \forall t_0 \in R, \tag{11b}$$

or equivalently

$$\int_{t_0}^{t} [h_B(t,\tau)h_A(\tau,t_0) - h_A(t,\tau)h_B(\tau,t_0)]d\tau = 0; \quad \forall t \geq t_0, \forall t_0 \in R. \tag{11c}$$

In fact, either of Eqs. (11b) or Eq. (11c) express necessary and sufficient conditions in terms of impulse responses for relaxed systems $A$ and $B$ to be commutative.

On the other hand, for commutativity of $B$ and $C$ under zero inintial conditions, the neccessary and suffifcient condition can be obtained from (11) by changing $A \to B$ and $B \to C$, the result is

$$h_{BC}(t,t_0) = h_{CB}(t,t_0); \quad \forall t \geq t_0, \forall t_0 \in R, \tag{12a}$$

$$\int_{t_0}^{t} h_C(t,\tau)h_B(\tau,t_0)d\tau = \int_{t_0}^{t} h_B(t,\tau)h_C(\tau,t_0)d\tau; \quad \forall t \geq t_0, \forall t_0 \in R, \tag{12b}$$

or equivalently,

$$\int_{t_0}^{t} [h_C(t,\tau)h_B(\tau,t_0) - h_B(t,\tau)h_C(\tau,t_0)]d\tau = 0; \quad \forall t \geq t_0, \forall t_0 \in R. \tag{12c}$$

For the proof of the commutativity of $C$ and $A$, that is the transitivity property of commutativity, it is required to show

$$h_{CA}(t,t_0) = h_{AC}(t,t_0); \quad \forall t \geq t_0, \forall t_0 \in R, \text{ that is,} \tag{13a}$$

$$\int_{t_0}^{t} h_A(t,\tau)h_C(\tau,t_0)d\tau = \int_{t_0}^{t} h_C(t,\tau)h_A(\tau,t_0)d\tau; \quad \forall t \geq t_0, \forall t_0 \in R, \tag{13b}$$

or equivalently,

$$\int_{t_0}^{t} [h_A(t,\tau)h_C(\tau,t_0) - h_C(t,\tau)h_A(\tau,t_0)]d\tau = 0; \quad \forall t \geq t_0, \forall t_0 \in R. \tag{13c}$$

In the light of the presence of equalities in (11b) and (12b), the work left for the proof of transitivity is to show the validity of (13b); if so then (13a) holds. This process will end with the proof of transitivity of linear time-varying relaxed systems of higher orders.

It worth's to remark that Eqs. (10a, b) are first used by J. E. Marshall to prove his assertion mentioned in Introduction (Eqs. (2, 3) in [2]). Later, they are used by the author for studying commutativity of Euler systems (Eqs. (34a, b) in [7]).

When we consider a scalar system

$$A: a_0(t)y_A(t) = x_A(t) \tag{14a}$$

for which its response $h_A(t,t_0)$ to a unit impulse $\delta(t - t_0)$ is obviously

$$h_A(t,t_0) = y_A(t) = \frac{1}{a_0(t)} x_A(t) = \frac{1}{a_0(t_0)} \delta(t - t_0). \tag{14b}$$

Then, let $B$ be a time-varying system of 1st order or higher with impulse response $h_B(t, t_0)$. Eqs. (10a) and (10b) lead to

$$h_{AB}(t, t_0) = \int_{t_0}^{t} h_B(t, \tau) \frac{\delta(\tau - t_0)}{a_0(t_0)} d\tau = \frac{h_B(t, t_0)}{a_0(t_0)}, \qquad (15a)$$

$$h_{BA}(t, t_0) = \int_{t_0}^{t} \frac{\delta(t - \tau)}{a_0(\tau)} h_B(\tau, t_0) d\tau = \frac{h_B(t, t_0)}{a_0(t)}. \qquad (15b)$$

From these equations, it is obvious that a scalar system that can be commutative with a 1st or higher order system must be a time-invariant system; that is $a_0(t) = a_0(t_0) = constant$ for all $t \geq t_0$. In other words, a scalar time-varying system cannot commutative with any order of higher order (1st, 2nd …) time-varying system. The only commutative pairs of it are the scalar systems. Though commutativity of a constant gain system (scalar time-invariant system) is always valid with time-varying system of any order, if the initial conditions exist the commutativity is possible only if the constant gain is 1; that is System $A$ is an identity.

The validation of transitivity using the equivalence of impulse responses $h_{AC}(t, t_0)$ and $h_{CA}(t, t_0)$ for first-order systems are considered in Section 5. But the extension of this result for systems at least one is 2nd order or higher is not straight forward and appears to be an unsolved problem. Instead of jumping this problem fully, we consider the special case that transitivity holds conditionally. Hence, the following discussion is not restricted by order limitation.

Let $A, B, C$ be linear time-varying systems with impulse responses $h_A, h_B, h_C$ satisfying

$$h_B(t, \tau) h_A(\tau, t_0) - h_A(t, \tau) h_B(\tau, t_0) = 0, \qquad (16a)$$

$$h_C(t, \tau) h_B(\tau, t_0) - h_B(t, \tau) h_C(\tau, t_0) = 0. \qquad (16b)$$

Eq. (16a) implies that Eqs. (11a, b, c) are satisfied, hence $(A, B)$ is a comutative pair. In a similar manner (16b) implies that $(B, C)$ is also a commutative pair.

Multiplying (16b) by $h_A(\tau, t_0)$ and inserting the equivalence of $h_B(t, \tau) h_A(\tau, t_0)$ as obtained from (16a), we find

$$h_C(t, \tau) h_B(\tau, t_0) h_A(\tau, t_0) - h_B(t, \tau) h_C(\tau, t_0) h_A(\tau, t_0) = 0, \qquad (17a)$$

$$h_C(t, \tau) h_B(\tau, t_0) h_A(\tau, t_0) - h_C(\tau, t_0) h_A(t, \tau) h_B(\tau, t_0) = 0, \qquad (17b)$$

$$[h_C(t,\tau)h_A(\tau,t_0) - h_A(t,\tau)h_C(\tau,t_0)]h_B(\tau,t_0) = 0. \tag{17c}$$

Since $h_B(\tau, t_0) \not\equiv 0$, it follows that

$$h_C(t,\tau)h_A(\tau,t_0) - h_A(t,\tau)h_C(\tau,t_0) = 0. \tag{18a}$$

This is sufficient to write

$$\int_{t_0}^{t} h_C(t,\tau)h_A(\tau,t_0)d\tau - \int_{t_0}^{t} h_A(t,\tau)h_C(\tau,t_0)d\tau = 0. \tag{18b}$$

Comparing with Eqs. (10a) and (10b), the first integral in (18b) is the impulse response of $AC$ and the second one is that of $CA$; hence, (18b) yields

$$h_{AC}(t,t_0) = h_{AC}(t,t_0). \tag{19}$$

So, that $A$ and $C$ are commutative. This proves the transitivity proper of higher-order systems under the set assumptions.

5. **Verifications for first-order systems**

This section verifies the transitivity property of first-order linear time-varying systems in their relaxed modes as presented in Section 3 by using the general results obtained for systems of any order in Section 4.

Consider a first-order linear time-varying system $A$ defined by

$$a_1(t)\frac{d}{dt}y_A(t) + a_0(t)y_A(t) = x_A(t), \tag{20}$$

with zero initial conditions. The impulse response $h_A(t, t_0)$ of this system to an impulse $\delta(t - t_0)$ occurring at $t = t_0$ is obviously expressed as [14, 15]

$$h_A(t,t_0) = \frac{1}{a_1(t_0)}e^{\int_{t_0}^{t} -\frac{a_0(\gamma)}{a_1(\gamma)}d\gamma} = \frac{1}{a_1(t_0)}e^{f_A(t) - f_A(t_0)}, \tag{21a}$$

where

$$f_A(\gamma) = \int -\frac{a_0(\gamma)}{a_1(\gamma)}d\gamma, \tag{21b}$$

$$\frac{d}{d\gamma}f_A(\gamma) = -\frac{a_0(\gamma)}{a_1(\gamma)}. \tag{21c}$$

Hence, for Systems $B$ and $C$ of the same type of $A$, it can be written that

$$h_B(t,t_0) = \frac{1}{b_1(t_0)} e^{\int_{t_0}^{t} -\frac{b_0(\gamma)}{b_1(\gamma)} d\gamma} = \frac{1}{b_1(t_0)} e^{f_B(t)-f_B(t_0)}, \qquad (22a)$$

$$f_B(\gamma) = \int -\frac{b_0(\gamma)}{b_1(\gamma)} d\gamma, \qquad (22b)$$

$$\frac{d}{d\gamma} f_B(\gamma) = -\frac{b_0(\gamma)}{b_1(\gamma)}, \qquad (22c)$$

$$h_C(t,t_0) = \frac{1}{c_1(t_0)} e^{\int_{t_0}^{t} -\frac{c_0(\gamma)}{c_1(\gamma)} d\gamma} = \frac{1}{c_1(t_0)} e^{f_C(t)-f_C(t_0)}, \qquad (23a)$$

$$f_C(\gamma) = \int_{t_0}^{t} -\frac{c_0(\gamma)}{c_1(\gamma)} d\gamma, \qquad (23b)$$

$$\frac{d}{d\gamma} f_C(\gamma) = -\frac{C_0(\gamma)}{C_1(\gamma)}. \qquad (23c)$$

For commutativity of $A$ and $B$, consider the verification of Eq. (11a) (or equivalently Eqs. (11b, c)). Using the relations in Eq. (2a) between the coefficients of $A$ and $B$ in Eq. (22b) and then recognizing (21b), we obtain

$$f_B(\gamma) = \int -\frac{b_0(\gamma)}{b_1(\gamma)} d\gamma = \int -\frac{k_1 a_0(\gamma) + k_0}{k_1 a_1(\gamma)} d\gamma = \int -\frac{a_0(\gamma)}{a_1(\gamma)} d\gamma - \frac{k_0}{k_1} \int \frac{1}{a_1(\gamma)} d\gamma$$

$$= f_A(\gamma) - \frac{k_0}{k_1} \int \frac{1}{a_1(\gamma)} d\gamma. \qquad (24a)$$

Using this result in Eq. (22a), we obtain

$$h_B(t,t_0) = \frac{1}{k_1 a_1(t_0)} e^{f_A(t) - \frac{k_0}{k_1} \int \frac{1}{a_1(t)} dt - f_A(t_0) + \frac{k_0}{k_1} \int \frac{1}{a_1(t_0)} dt_0},$$

$$= \frac{1}{k_1} \frac{1}{a_1(t_0)} e^{f_A(t) - f_A(t_0)} e^{g(t_0) - g(t)} = \frac{1}{k_1} e^{g(t_0) - g(t)} h_A(t,t_0), \qquad (24b)$$

where

$$g(t) = \frac{k_0}{k_1} \int \frac{1}{a_1(t)} dt. \qquad (24c)$$

By using (15b) in (7a), we obtain

$$h_{AB}(t,t_0) = \int_{t_0}^{t} \frac{1}{k_1} e^{g(\tau)-g(t)} \; h_A(t,\tau) h_A(\tau,t_0) d\tau. \tag{25}$$

Before proceeding further, we show the transitivity property of $h_A(\bullet,\bullet)$; in fact using (21a)

$$h_A(t,\tau) h_A(\tau,t_0) = \frac{1}{a_1(\tau)} e^{f_A(t)-f_A(\tau)} \frac{1}{a_1(t_0)} e^{f_A(\tau)-f_A(t_0)}$$

$$= \frac{1}{a_1(\tau)} \frac{1}{a_1(t_0)} e^{f_A(t)-f_A(t_0)} = \frac{1}{a_1(\tau)} h_A(t,t_0). \tag{26}$$

Same property holds for the remaining first-order systems $B$ and $C$ as well. Another relation we need in the future proves is arrived as follows:

$$\int_{t_0}^{t} \frac{e^{g(\tau)}}{a_1(\tau)} d\tau = \int_{t_0}^{t} \frac{e^{\frac{k_0}{k_1}\int \frac{1}{a_1(\tau)}d\tau}}{a_1(\tau)} d\tau = \frac{k_1}{k_0} e^{\frac{k_0}{k_1}\int \frac{1}{a_1(\tau)}d\tau} \bigg|_{t_0}^{t}$$

$$= \frac{k_1}{k_0} e^{g(\tau)} \bigg|_{t_0}^{t} = \frac{k_1}{k_0} \left[ e^{g(t)} - e^{g(t_0)} \right]. \tag{27a}$$

Similarly, we obtain

$$\int_{t_0}^{t} \frac{e^{-g(\tau)}}{a_1(\tau)} d\tau = \frac{k_1}{k_0} \left[ e^{-g(t_0)} - e^{-g(t)} \right]. \tag{27b}$$

Using (26) and (27a) in (25), we have

$$h_{AB}(t,t_0) = \int_{t_0}^{t} \frac{1}{k_1} e^{g(\tau)-g(t)} \frac{1}{a_1(\tau)} h_A(t,t_0) d\tau = \frac{e^{-g(t)}}{k_1} h_A(t,t_0) \int_{t_0}^{t} \frac{e^{g(\tau)}}{a_1(\tau)} d\tau$$

$$= \frac{e^{-g(t)}}{k_1} h_A(t,t_0) \frac{k_1}{k_0} \left[ e^{g(t)} - e^{g(t_0)} \right] = \frac{1}{k_0} h_A(t,t_0) \left[ 1 - e^{g(t_0)-g(t)} \right]. \tag{28}$$

Using (24b) in (10b) and using the formula (26) and (27b), we obtain sequentially

$$h_{BA}(t,t_0) = \int_{t_0}^{t} h_A(t,\tau) \frac{1}{k_1} e^{g(t_0)-g(\tau)} \; h_A(\tau,t_0) d\tau$$

$$= \int_{t_0}^{t} \frac{1}{k_1} e^{g(t_0)-g(\tau)} \frac{1}{a_1(\tau)} h_A(t,t_0) d\tau$$

$$= \frac{e^{g(t_0)}}{k_1} h_A(t,t_0) \int_{t_0}^{t} \frac{e^{-g(\tau)}}{a_1(\tau)} d\tau = \frac{e^{g(t_0)}}{k_1} h_A(t,t_0) \frac{k_1}{k_0} \left[ e^{-g(t_0)} - e^{-g(t)} \right]$$

$$= \frac{1}{k_0} h_A(t,t_0) \left[ 1 - e^{g(t_0)-g(t)} \right]. \tag{29}$$

Obviously, $AB$ and $BA$ have the same impulse response as seen in Eqs. (28) and (29), respectively; so $A$ and $B$ are commutative.

**Remark 1**: Although Eq. (11c), a property of impulse response of the commutative systems $A$ and $B$, seems to mislead to the conclusion of the integrant being zero; that is,

$$\Delta(t_0, \tau, t) \triangleq h_B(t,\tau) h_A(\tau, t_0) - h_A(t,\tau) h_B(\tau, t_0) = 0. \tag{30}$$

This is strictly wrong. In fact, using (24b) for $h_A(t, t_0)$, and (26) later, we have

$$\Delta(t_0, \tau, t) = \frac{1}{k_1} e^{g(\tau)-g(t)} h_A(t,\tau) h_A(\tau,t_0) - h_A(t,\tau) \frac{1}{k_1} e^{g(t_0)-g(\tau)} h_A(\tau, t_0) = 0$$

$$= \frac{1}{k_1} h_A(t,\tau) h_A(\tau, t_0) \left[ e^{g(\tau)-g(t)} - e^{g(t_0)-g(\tau)} \right]$$

$$= \frac{1}{k_1} h_A(t, t_0) \frac{\left[ e^{g(\tau)-g(t)} - e^{g(t_0)-g(\tau)} \right]}{a_1(\tau)} = 0. \tag{31a}$$

Since, $h_A(t, t_0) \not\equiv 0$ and $a_1(\tau) \not\equiv 0$, this implies

$$e^{g(\tau)-g(t)} = e^{g(t_0)-g(\tau)}, \tag{31b}$$

or equivalently

$$e^{2g(\tau)} = e^{g(t)+g(t_0)}; \quad g(\tau) = \frac{1}{2}[g(t) + g(t_0)]. \tag{31c}$$

With (24c), the last equation is equivalent to

$$\int \frac{1}{a_1(\tau)} d\tau = \frac{1}{2} \left[ \int \frac{1}{a_1(t)} dt + \int \frac{1}{a_1(t_0)} d t_0 \right], \tag{31d}$$

which is impossible for an arbitrary function $a_1(\bullet)$ And for general values of $t_0 \leq \tau \leq t$.

Similar result obtained in Eqs. (25-29) can be obtained for commutative pair $(B, C)$; either by the parallel derivations made for $(A, B)$, or by replacing $A \to B, B \to C$, so $a_1 \to b_1$ and $k_1, k_0$ by $\ell_1, \ell_0$, respectively.

For the commutativity of $A$ and $C$, to show that $h_{AC}(t, t_0) = h_{CA}(t, t_0)$, the replacements $B \to C, (k_1, k_0) \to (k_1 \ell_1, k_0 \ell_1 + \ell_0)$ in Eqs. (25-29) complete the verification of transitivity for initially relaxed first-order linear time-varying systems by using impulse responses.

## 6. Example

This example is given for illustration of the results presented in Sections 4 and 5. Consider system $A$ describe by

$$(t+1)\dot{y}_A + (t+2)y_A = x_A, \quad -1 < t_0 \le t, \tag{32}$$

where $a_1(t) = t+1$, $a_0(t) = t+2$; $f(t)$ for $A$ is computed by using (21b) as

$$f_A(t) = \int -\frac{t+2}{t+1} dt = -t - \ln(t+1). \tag{33}$$

Then, by (21a), its impulse response is computed as

$$h_A(t, t_0) = \frac{e^{-t-\ln(t+1)+t_0+\ln(t_0+1)}}{t_0+1} = \frac{e^{-t+t_0+\ln\frac{t_0+1}{t+1}}}{t_0+1} = \frac{1}{t+1} e^{-t+t_0}. \tag{34}$$

Let subsystem $B$ be described by

$$2(t+1)\dot{y}_B + (2t+5)y_B; \quad -1 < t_0 \le t, \tag{35}$$

where $b_1(t) = 2(t+1)$, $b_0(t) = 2t+5$ which can be obtained from $a_1$ and $a_0$ by using Eq. (2a) with $k_1 = 2$, $k_0 = 1$; hence, $(A, B)$ is a commutative pair. Using (22b), $f(t)$ for $B$ can be obtained as

$$f_B(t) = \int -\frac{2t+5}{2(t+1)} dt = -t - 1.5\ln(t+1). \tag{36}$$

Then, by (22a), the impulse response of $B$ is computed as

$$h_B(t, t_0) = \frac{e^{-t-1.5\ln(t+1)+t_0+1.5\ln(t_0+1)}}{2(t_0+1)} = \frac{e^{-t+t_0+1.5\ln\frac{t_0+1}{t+1}}}{2(t_0+1)}$$

$$= \frac{1}{2(t_0 + 1)} e^{-t+t_0} \frac{(t_0 + 1)^{1.5}}{(t + 1)^{1.5}} = \frac{(t_0 + 1)^{0.5}}{2(t + 1)^{1.5}} e^{-t+t_0}. \tag{37}$$

For computations of $h_{AB}(t, t_0)$ and $h_{BA}(t, t_0)$, first consider the integrands in Eqs. (10a), (10b) and (11b).

$$h_B(t, \tau)h_A(\tau, t_0) = \frac{(\tau + 1)^{0.5}}{2(t + 1)^{1.5}} e^{-t+\tau} \frac{1}{\tau + 1} e^{-\tau+t_0} = \frac{1}{2(t + 1)^{1.5}(\tau + 1)^{0.5}} e^{-t+t_0}, \tag{38a}$$

$$h_A(t, \tau)h_B(\tau, t_0) = \frac{1}{t+1} e^{-t+\tau} \frac{(t_0+1)^{0.5}}{2(\tau+1)^{1.5}} e^{-\tau+t_0} = \frac{(t_0+1)^{0.5}}{2(t+1)(\tau+1)^{1.5}} e^{-t+t_0}. \tag{38b}$$

Obviously, these integrands are not identical and Eq. (30) in connection to Remark 1 is not satisfied for the general values of $t_0 \leq \tau \leq t \in R$. The impulse response of $AB$ and $BA$ are computed as in (10a) and (10b), that is the integral of the integrands in (38a) and (38b), respectively. The results are as follows:

$$h_{AB}(t, t_0) = \int_{t_0}^{t} h_B(t, \tau)h_A(\tau, t_0) d\tau = \int_{t_0}^{t} \frac{e^{-t+t_0}}{2(t + 1)^{1.5}} \frac{1}{(\tau + 1)^{0.5}} d\tau$$

$$= \frac{e^{-t+t_0}}{2(t + 1)^{1.5}} \frac{(\tau + 1)^{0.5}}{0.5} \Big|_{t_0}^{t}$$

$$= \frac{e^{-t+t_0}}{(t + 1)^{1.5}} [(t + 1)^{0.5} - (t_0 + 1)^{0.5}] = \frac{e^{-t+t_0}}{t + 1} \left[1 - \left(\frac{t_0 + 1}{t + 1}\right)^{0.5}\right], \tag{39a}$$

$$h_{BA}(t, t_0) = \int_{t_0}^{t} \frac{(t_0 + 1)^{0.5} e^{-t+t_0}}{2(t + 1)} \frac{1}{(\tau + 1)^{1.5}} d\tau = \frac{(t_0 + 1)^{0.5} e^{-t+t_0}}{2(t + 1)} \frac{(\tau + 1)^{-0.5}}{-0.5} \Big|_{t_0}^{t}$$

$$= \frac{(t_0+1)^{0.5} e^{-t+t_0}}{(t+1)} \left[\frac{1}{(t_0+1)^{0.5}} - \frac{1}{(t+1)^{0.5}}\right] = \frac{e^{-t+t_0}}{t+1} \left[1 - \left(\frac{t_0+1}{t+1}\right)^{0.5}\right]. \tag{39b}$$

Obviously, $h_{AB}(t, t_0) \equiv h_{BA}(t, t_0)$.

When System $C$ is obtained from $B$ by (2b) with $\ell_1 = -0.5, \ell_0 = 3.5$, its coefficients are found as $c_1 = -(t + 1), c_0 = -(2t + 5)(-0.5) + 3.5 = -t + 1$. Then, $f(t)$ for $C$ is computed from (23b) as

$$f_C(t) = \int -\frac{-t + 1}{-t - 1} dt = -t + 2\ln(t + 1). \tag{40a}$$

Using (23a), the impulse response of $C$ is now obtained as

$$h_C(t, t_0) = \frac{e^{-t+2\ln(t+1)+t_0-2\ln(t_0+1)}}{-(t_0+1)} = -\frac{1}{t_0+1} e^{-t+t_0} \left(\frac{t+1}{t_0+1}\right)^2$$

$$= -\frac{(t+1)^2}{(t_0+1)^3} e^{-t+t_0}. \tag{40b}$$

The impulse response of the connection $AC$ is computed by Eq. (10a) modified by replacing $B$ by $C$, and using Eqs. (34) and (40b) for $A$ and $C$; we obtain

$$h_{AC}(t, t_0) = \int_{t_0}^{t} h_C(t, \tau) h_A(\tau, t_0) d\tau = \int_{t_0}^{t} -\frac{(t+1)^2}{(\tau+1)^3} e^{-t+\tau} \frac{1}{\tau+1} e^{-\tau+t_0} d\tau$$

$$= -(t+1)^2 e^{-t+t_0} \int_{t_0}^{t} \frac{1}{(\tau+1)^4} d\tau = \frac{(t+1)^2 e^{-t+t_0}}{3} \frac{1}{(\tau+1)^3} \Big|_{t_0}^{t}$$

$$= \frac{(t+1)^2 e^{-t+t_0}}{3} \left[\frac{1}{(t+1)^3} - \frac{1}{(t_0+1)^3}\right]$$

$$= \frac{e^{-t+t_0}}{3(t+1)} \left[1 - \left(\frac{t+1}{t_0+1}\right)^3\right]. \tag{41a}$$

Note that the impulse response of the connection $CA$ can be obtained similarly by using Eq. (10a) by replacing $B$ by $C$ and using Eqs. (40b) and (34) for $C$ and $A$; the result is

$$h_{CA}(t, t_0) = \int_{t_0}^{t} h_A(t, \tau) h_C(\tau, t_0) d\tau = -\int_{t_0}^{t} \frac{1}{t+1} e^{-t+\tau} \frac{(\tau+1)^2}{(t_0+1)^3} e^{-\tau+t_0} d\tau$$

$$= -\frac{e^{-t+t_0}}{(t+1)(t_0+1)^3} \int_{t_0}^{t} (\tau+1)^2 d\tau = -\frac{e^{-t+t_0}}{3(t+1)(t_0+1)^3} (\tau+1)^3 \Big|_{t_0}^{t}$$

$$= \frac{e^{-t+t_0}}{3(t+1)(t_0+1)^3} [(t_0+1)^3 - (t+1)^3]$$

$$= \frac{e^{-t+t_0}}{3(t+1)} \left[1 - \left(\frac{t+1}{t_0+1}\right)^3\right]. \tag{41b}$$

This is the same impulse response expressed in (41a). Hence, it is verified that $(A, C)$ is a commutative pair.

7. **Transitivity for Some 1$^{st}$ and 2$^{nd}$ Order Relaxed Systems**

Let $A, B, C$ be linear time-varying systems of orders 2, 1, 1; respectively. Let they be represented by the general form of (1) with coefficients $a_2, a_1, a_0$; $b_1, b_0$; $c_1, c_0$; inputs by $x_A$, $x_B$, $x_C$; output by $y_A$, $y_B$, $y_C$. It is true that $A$ has 2$^{nd}$ or lower order commutative pairs $B$ if and only if [8-10]

$$-a_2^{0.5} \frac{d}{dt}\left[a_0 - \frac{4a_1^2 + 3\dot{a}_2^2 - 8a_1\dot{a}_2 + 8\dot{a}_1 a_2 - 4a_2\ddot{a}_2}{16 a_2}\right] k_1 = 0, \qquad (42a)$$

$$\begin{bmatrix} b_2 \\ b_1 \\ b_0 \end{bmatrix} = \begin{bmatrix} a_2 & 0 & 0 \\ a_1 & a_2^{0.5} & 0 \\ a_0 & \dfrac{a_2^{-0.5}(2a_1 - \dot{a}_2)}{4} & 1 \end{bmatrix} \begin{bmatrix} k_2 \\ k_1 \\ k_0 \end{bmatrix}, \qquad (42b)$$

where $k_2, k_1, k_0$ are arbitrary constants. We assume that $B$ is a finite order commutative pair of $A$; hence $k_2 = 0$ and

$$\begin{bmatrix} b_1 \\ b_0 \end{bmatrix} = \begin{bmatrix} a_2^{0.5} & 0 \\ \dfrac{a_2^{-0.5}(2a_1 - \dot{a}_2)}{4} & 1 \end{bmatrix} \begin{bmatrix} k_1 \\ k_0 \end{bmatrix}. \qquad (43)$$

On the other hand, it is true that $B$ has 1$^{st}$ or lower order commutative pairs $C$, if [8]

$$\begin{bmatrix} c_1 \\ c_0 \end{bmatrix} = \begin{bmatrix} b_1 & 0 \\ b_0 & 1 \end{bmatrix} \begin{bmatrix} \ell_1 \\ \ell_0 \end{bmatrix}, \qquad (44)$$

where $\ell_1$ and $\ell_0$ are arbitrary constants. Inserting the values of $b_1$ and $b_0$ from Eq. (43),) into Eq. (44), and rearranging, we obtain

$$\begin{bmatrix} c_1 \\ c_0 \end{bmatrix} = \begin{bmatrix} a_2^{0.5} & 0 \\ \dfrac{a_2^{-0.5}(2a_1 - \dot{a}_2)}{4} & 1 \end{bmatrix} \begin{bmatrix} k_1 \ell_1 \\ k_0 \ell_1 + \ell_0 \end{bmatrix}. \qquad (45)$$

It is obviously true that this equation is in the form of (43). So $C$ is among the first-order commutative pairs of $A$. Hence, it has been proven that if $(A, B)$ and $(B, C)$ are commutative pairs so is $(A, C)$. Hence, transitivity is valid.

Let's get one step up and consider transitivity property between two second order and one first-order systems. Let $A, B, C$ be 2$^{nd}$, 1$^{st}$, 2$^{nd}$ order systems, respectively, with coefficient $a_2, a_1, a_0$; $b_1, b_0, c_2, c_1, c_0$. Assume $(A, B)$ and $(B, C)$ are commutative pairs and show $(A, C)$ is also commutative. Eq. (43) is valid since $(A, B)$ is a commutative pair. Similar equation

$$\begin{bmatrix} b_1 \\ b_0 \end{bmatrix} = \begin{bmatrix} c_2^{0.5} & 0 \\ \dfrac{c_2^{-0.5}(2c_1 - \dot{c}_2)}{4} & 1 \end{bmatrix} \begin{bmatrix} \ell_1 \\ \ell_0 \end{bmatrix} \qquad (46)$$

is also valid since $(B, C)$ is a commutative pair. Further, (42a) is valid and the similar equation for $C$ is written as

$$-c_2^{0.5} \dfrac{d}{dt}\left[c_0 - \dfrac{4c_1^2 + 3\dot{c}_2^2 - 8c_1\dot{c}_2 + 8\dot{c}_1 c_2 - 4c_2\ddot{c}_2}{16 c_2}\right]\ell_1 = 0. \qquad (47)$$

Since $B$ is of order 1, $k_1$ in (43) and $\ell_1$ in (46) are not zero; hence (42a) and (47) yield

$$a_0 = \left[A_0 + \dfrac{4a_1^2 + 3\dot{a}_2^2 - 8a_1\dot{a}_2 + 8\dot{a}_1 a_2 - 4a_2\ddot{a}_2}{16 a_2}\right], \qquad (48a)$$

$$c_0 = \left[C_0 + \dfrac{4c_1^2 + 3\dot{c}_2^2 - 8c_1\dot{c}_2 + 8\dot{c}_1 c_2 - 4c_2\ddot{c}_2}{16 c_2}\right], \qquad (48b)$$

respectively, where $A_0$ and $C_0$ are arbitrary constants.

From Eq. (43), the first line, we obtain

$$a_2 = \dfrac{b_1^2}{k_1^2} \;\to\; \dot{a}_2 = \dfrac{2 b_1 \dot{b}_1}{k_1^2}, \qquad \ddot{a}_2 = \dfrac{2(\dot{b}_1^2 + b_1 \ddot{b}_1)}{k_1^2}. \qquad (49a)$$

Using these results in Eq. (43), the second line, we obtain

$$a_1 = \dfrac{b_1(2 b_0 - 2 k_0 + \dot{b}_1)}{k_1^2} \;\to\; \dot{a}_1 = \dfrac{\dot{b}_1(2 b_0 - 2 k_0 + \dot{b}_1) + b_1(2\dot{b}_0 + \ddot{b}_1)}{k_1^2}. \qquad (49b)$$

Inserting the values of $a_2, \dot{a}_2, \ddot{a}_2, a_1, \dot{a}_1$ in Eq. (49) into Eq. (48a) and simplifying, we obtain

$$a_0 = A_0 + \left[\dfrac{b_0 - k_0}{k_1}\right]^2 + \dfrac{b_1 \dot{b}_0}{k_1^2}. \qquad (49c)$$

Similar results can be obtained starting from (45) and using (48b):

$$c_2 = \dfrac{b_1^2}{\ell_1^2}, \qquad (50a)$$

$$c_1 = \dfrac{b_1(2 b_0 - 2\ell_0 + \dot{b}_1)}{\ell_1^2}, \qquad (50b)$$

$$c_0 = C_0 + \left[\frac{b_0 - \ell_0}{\ell_1}\right]^2 + \frac{b_1 \dot{b}_0}{\ell_1^2}. \tag{50c}$$

To prove the commutativity of $A$ and $C$, it is sufficient to show that

$$\begin{bmatrix} c_2 \\ c_1 \\ c_0 \end{bmatrix} = \begin{bmatrix} a_2 & 0 & 0 \\ a_1 & a_2^{0.5} & 0 \\ a_0 & \dfrac{a_2^{-0.5}(2a_1 - \dot{a}_2)}{4} & 1 \end{bmatrix} \begin{bmatrix} m_2 \\ m_1 \\ m_0 \end{bmatrix}, \tag{51}$$

which is similar equation to (42b), where $m_2, m_1, m_0$ are some constants. It is true that, the coefficients of $C$ can be expressed as in (51) by proper choice of constants $m_2, m_1, m_0$; in fact, inserting in coefficients $a_i$ in Eq. (49) and coefficient $c_i$ in Eq. (50) into Eq. (51), we find that

$$m_2 = \frac{k_1^2}{\ell_1^2}, \tag{52a}$$

$$m_1 = \frac{2k_1(k_0 - \ell_0)}{\ell_1^2}, \tag{52b}$$

$$m_0 = C_0 - A_0 \frac{k_1^2}{\ell_1^2} + \left[\frac{k_0 - \ell_0}{\ell_1}\right]^2, \tag{52c}$$

which are all constants. Hence, Eq. (51) together with Eq. (42) imply that $C$ is a second-order commutative pair of $A$, that is $A$ and $C$ are also commutative. This proves the transitivity property of $(A, B)$ and $(B, C)$ to $(A, C)$.

We express the result by a theorem:

***Theorem 3:*** *The commutativity property between three subsystems of which at most 2 of them second order and the other(s) are first-order always satisfy the transitivity property under zero initial conditions.*

## 8. Conclusions

Transitivity property of commutativity is defined for commutative linear time-varying systems of any order. The transitivity results are presented for first-order systems with and without initial conditions. The study is carried on two approaches; one is based on use of impulse response function and the other depends on use of general conditions set by the author and others for commutativity of linear time-varying systems.

One illustrative example is included to show the validity of the results. However, general transitivity property of commutative high order systems have not been proved or disproved and this remains as an unsolved problem yet.

**REFERENCES**

[1]  R. Boylestad and L. Nashelsky, Electronic Devices and Circuit Theory, Prentice Hall, New Jersey, 2013.

[2]  E. Marshal, Commutativity of time varying systems, Electronics Letters, 13 (1977) 539-540.

[3]  M. Koksal, Commutativity of second order time-varying systems, International Journal of Control, 36 (1982) 541-544.

[4]  S. V. Salehi, Comments on 'Commutativity of second-order time-varying systems', International Journal of Control, 37 (1983) 1195-1196.

[5]  M. Koksal, Corrections on 'Commutativity of second-order time-varying systems', International Journal of Control, 38 (1983) 273-274.

[6]  M. Koksal, General conditions for the commutativity of time-varying systems, IASTED International Conference on Telecommunication and Control (TELCONi84), 1984, pp. 223-225.

[7]  M. Koksal, A Survey on the Commutativity of time-varying systems, METU, Gaziantep Engineering Faculty, Technical. Report no: GEEE CAS-85/1, 1985.

[8]  M. Koksal, An exhaustive study on the commutativity of time-varying systems, International Journal of Control, 47 (1988) 1521-1537.

[9]  M. Koksal and M. E. Koksal, Commutativity of linear time-varying differential systems with non-zero initial conditions: A review and some new extensions, Mathematical Problems in Engineering, 2011 (2011) 1-25.

[10] M. E. Koksal and M. Koksal, Commutativity of cascade connected discrete time linear time-varying systems, 2013 Automatic Control National Meeting TOK'2013, (2013) p.1128-1131.

[11] M. Koksal and M. E. Koksal, Commutativity of cascade connected discrete-time linear time-varying systems, Transactions of the Institute of Measurement and Control, 37 (2015) 615-622.

[12] M. E. Koksal, The Second order commutative pairs of a first-order linear time-varying system, Applied Mathematics and Information Sciences, 9 (2015) 1-6.

[13] M. E. Koksal, Decomposition of a second-order linear time-varying differential system

as the series connection of two first-order commutative pairs, Open Mathematics, 14 (2016) 693-704.

[14]  C. A. Desoer, Notes For A Second Course On Linear Systems, Van Nostrand Rheinhold, New York, 1970.

[15]  Chi-Tsong Chen, Linear System Theory and Design, New York Oxford, Oxford University Press, 1999.